\documentclass[%
 reprint,
superscriptaddress,
 amsmath,amssymb,
 aps,
prb,
]{revtex4-2}

\usepackage{graphicx}

\usepackage{appendix}

\usepackage[normalem]{ulem}

\usepackage{dcolumn}
\usepackage{bm}
\usepackage{float}
\usepackage{xcolor}
\usepackage[colorlinks = true,
            linkcolor = blue,
            urlcolor  = blue,
            citecolor = blue,
            anchorcolor = blue]{hyperref}

\usepackage{orcidlink}

\newcommand{\orcid}[1]{\href{https://orcid.org/#1}{\textcolor[HTML]{A6CE39}{\aiOrcid}}}

\begin{document}

\preprint{APS/123-QED}

\title{Strain induced magnetic phase transitions in $\mathrm{Fe_{3}GeTe_{2}}$ monolayer}

\author{Anjali Jyothi Bhasu\,\orcidlink{0009-0001-9433-0822}}
\affiliation{
Department of Theoretical Physics, Institute of Physics,
\href{https://ror.org/02w42ss30}{Budapest University of Technology and Economics},
M\H{u}egyetem rkp.~3., HU-1111 Budapest, Hungary
}

\author{Satish Kumar}%
\affiliation{Government Degree College Bilaspur, Bilaspur-174001, Himachal Pradesh, India}

\author{M\'aty\'as T\"or\"ok}%
\affiliation{
Department of Theoretical Physics, Institute of Physics,
\href{https://ror.org/02w42ss30}{Budapest University of Technology and Economics},
M\H{u}egyetem rkp.~3., HU-1111 Budapest, Hungary
}

\author{D\'aniel Tibor Pozs\'ar\,\orcidlink{0009-0009-6169-9066}}%
\affiliation{Department of Physics of Complex Systems, \href{https://ror.org/01jsq2704}{E$\ddot{o}$tv$\ddot{o}$s Lor\'and University},  1117 Budapest, Hungary}

\author{Bendeg\'uz Ny\'ari\,\orcidlink{0000-0001-5524-9995}}%
\affiliation{
Department of Theoretical Physics, Institute of Physics,
\href{https://ror.org/02w42ss30}{Budapest University of Technology and Economics},
M\H{u}egyetem rkp.~3., HU-1111 Budapest, Hungary
}
\affiliation{HUN-REN-BME Condensed Matter Research Group,
\href{https://ror.org/02w42ss30}{Budapest University of Technology and Economics}, M\H{u}egyetem rkp.~3., HU-1111 Budapest, Hungary}

\author{L\'aszl\'o Udvardi\,\orcidlink{0000-0002-0822-2439}}%
\affiliation{
Department of Theoretical Physics, Institute of Physics,
\href{https://ror.org/02w42ss30}{Budapest University of Technology and Economics},
M\H{u}egyetem rkp.~3., HU-1111 Budapest, Hungary
}

\author{Gabriel Mart\'inez-Carracedo\,\orcidlink{0000-0003-4351-897X} }%
\affiliation{Departamento de Física, \href{https://ror.org/006gksa02} {Universidad de Oviedo}, 33007 Oviedo, Spain}
\affiliation{\href{https://ror.org/03ppnws78}{Centro de Investigación en Nanomateriales y Nanotecnología, Universidad de Oviedo}-CSIC, 33940 El Entrego, Spain}

\author{Bal\'azs Nagyfalusi\,\orcidlink{0000-0003-3875-2059}}%
\affiliation{Departamento de Física, \href{https://ror.org/006gksa02} {Universidad de Oviedo}, 33007 Oviedo, Spain}

\author{Amador Garc\'ia-Fuente\,\orcidlink{0000-0002-4570-8315}}%
\affiliation{Departamento de Física, \href{https://ror.org/006gksa02} {Universidad de Oviedo}, 33007 Oviedo, Spain}

\author{Jaime Ferrer\,\orcidlink{0000-0002-4067-2325}}%
\affiliation{Departamento de Física, \href{https://ror.org/006gksa02} {Universidad de Oviedo}, 33007 Oviedo, Spain}
\author{Zolt\'{a}n Tajkov\,\orcidlink{0000-0001-9733-0361}}
\affiliation{\href{https://ror.org/05wswj918}{HUN-REN Centre for Energy Research}, Institute of Technical Physics and Materials Science, Konkoly-Thege Miklós út 29-33., 1121 Budapest, Hungary}
\affiliation{Department of Physics of Complex Systems, \href{https://ror.org/01jsq2704}{E$\ddot{o}$tv$\ddot{o}$s Lor\'and University},  1117 Budapest, Hungary}

\author{L\'aszl\'o Oroszl\'any\,\orcidlink{0000-0001-5682-6424}}%
\affiliation{Department of Physics of Complex Systems, \href{https://ror.org/01jsq2704}{E$\ddot{o}$tv$\ddot{o}$s Lor\'and University},  1117 Budapest, Hungary}
\affiliation{
\href{https://ror.org/035dsb084}{HUN-REN Wigner Research Centre for Physics},Konkoly-Thege Miklós út 29-33., 1121 Budapest, Hungary
}

\author{Levente R\'ozsa\,\orcidlink{0000-0001-9456-5755}}%
\affiliation{
\href{https://ror.org/035dsb084}{HUN-REN Wigner Research Centre for Physics},Konkoly-Thege Miklós út 29-33., 1121 Budapest, Hungary
}
\affiliation{
Department of Theoretical Physics, Institute of Physics,
\href{https://ror.org/02w42ss30}{Budapest University of Technology and Economics},
M\H{u}egyetem rkp.~3., HU-1111 Budapest, Hungary
}

\author{L\'aszl\'o Szunyogh\,\orcidlink{0000-0001-7430-3627}}%
\affiliation{
Department of Theoretical Physics, Institute of Physics,
\href{https://ror.org/02w42ss30}{Budapest University of Technology and Economics},
M\H{u}egyetem rkp.~3., HU-1111 Budapest, Hungary
}
\affiliation{HUN-REN-BME Condensed Matter Research Group,
\href{https://ror.org/02w42ss30}{Budapest University of Technology and Economics}, M\H{u}egyetem rkp.~3., HU-1111 Budapest, Hungary}


\begin{abstract}
We investigate the magnetic properties of a monolayer of $\mathrm{Fe_{3}GeTe_{2}}$ as a function of the lattice constant by combining first-principles calculations with atomistic spin dynamics simulations. The calculated magnetic exchange interactions reveal a competition between ferromagnetic and antiferromagnetic couplings, with the latter being significantly strengthened under compressive strain. Stochastic Landau–Lifshitz–Gilbert simulations reveal a substantial decrease in the Curie temperature with decreasing lattice constant, and predict a transition of the magnetic ground state from a ferromagnetic configuration to a conical spin-spiral state. We introduce a simple spin-model which explains the stabilization of the spiral phase due to competing exchange interactions. We found multiple magnetic phase transitions involving ferromagnetic, conical spin-spiral, and planar Néel states, depending on both the lattice constant and the temperature. The absence of Dzyaloshinskii–Moriya interactions is found to significantly reduce the Néel temperature, while leaving the Curie temperature largely unaffected. Our findings reveal the importance of lattice distortions in controlling complex magnetic phases and their evolution with temperature.
\end{abstract}

\maketitle


\begin{figure*}
    \includegraphics[width=\linewidth]{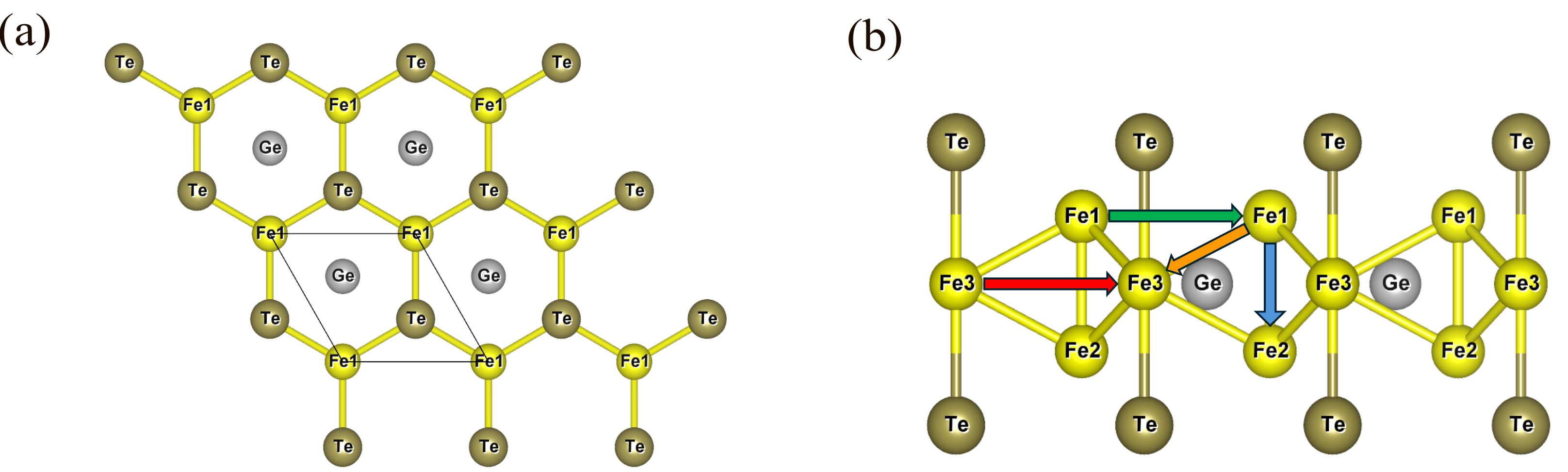}
    \caption{(a) Top and (b) side view of a monolayer $\mathrm{Fe_{3}GeTe_{2}}$. The unit cell is indicated by thin black lines. Fe1, Fe2 and Fe3 denote the Fe sites being inequivalent by 2D translational symmetry. Note that the sites Fe1 and Fe2 are equivalent by mirror symmetry. In panel (b), the blue, orange, green and red arrows indicate the first-, second-, and two different third-nearest-neighbor interactions between the Fe sites, respectively. }
    \label{fig:FGT}
\end{figure*}

\section{\label{sec:intro}Introduction}
The discovery of intrinsic long-range magnetic order in atomically thin materials has sparked significant interest in two-dimensional (2D) magnetism \cite{huang2017layer,gong2017discovery}. In 2D systems, thermal spin fluctuations suppress long-range order as stated by the Mermin-Wagner theorem \cite{mermin1966absence}. However, magnetic anisotropy can stabilize magnetic ordering by introducing an energy gap in the excitation spectrum, leading to a finite Curie temperature, $T_\mathrm{C} $ \cite{huang2017layer,xu2018interplay}. The high degree of control over the microscopic parameters make 2D magnets attractive candidates for spintronic devices, especially magnetic memory applications, while the capability to exfoliate them further enhances their technological potential \cite{yang2022two, fert2024electrical}.

Among these materials, Fe$_n$GeTe$_2$ compounds ($n=3,4,5$), commonly referred to as FGT systems, have attracted considerable attention due to their metallic character and relatively high Curie temperatures compared to other 2D magnets. $\mathrm{Fe_{3}GeTe_{2}}$ crystallizes in a hexagonal structure with $\mathrm{P6_{3}/mmc}$ space-group symmetry, where $\mathrm{Fe_{3}Ge}$ layers are separated by two Te layers that are coupled by van der Waals forces, see Fig. \ref{fig:FGT}. Experimentally, a monolayer of $\mathrm{Fe_{3}GeTe_{2}}$ exhibits a $T_\mathrm{C} $ of approximately 130~K \cite{fei2018two}. External perturbations have been shown to significantly influence their magnetic properties. Under hydrostatic pressure it has been observed that the ferromagnetic state of $\mathrm{Fe_{3}GeTe_{2}}$ vanishes above a certain pressure, resulting in an intermediate labyrinthine-domain state \cite{HeshenWang}. The decrease in  $T_\mathrm{C} $ with increasing pressure is attributed to an increase in the ratio of the exchange interaction to the magnetocrystalline anisotropy. 
Additionally, pressure-dependent studies reveal a critical modulation of the anomalous Hall effect \cite{XiangqiWang}, while a compressive strain of 0.4~\% in $\mathrm{Fe_{3}GeTe_{2}}$/$\mathrm{\alpha-In_{2}Se_{3}}$ heterostructures reduces $T_\mathrm{C}$ by 20~K \cite{RyujiFujita}. Complex spin textures, including Bloch-type skyrmions and stripe domains, have also been observed near $T_\mathrm{C}$ \cite{ding2022tuning}. Experiments also suggest that bulk and bilayer $\mathrm{Fe_{3}GeTe_{2}}$ may exhibit competing ferromagnetic (FM) and antiferromagnetic (AFM) phases, challenging earlier interpretations of ferromagnetic ground states \cite{Kim_2019,Yi_2017}.

First-principles calculations provided important insights into the magnetic behavior of FGT systems. 
Monolayer $\mathrm{Fe_{3}GeTe_{2}}$ has been predicted to exhibit a ferromagnetic ground state \cite{Zhen-Xiong,ghosh2023unraveling}, with tensile strain further stabilizing the ferromagnetic order \cite{PUSHKAREV2023171456}. However, density-functional theory (DFT) and DFT+U approaches are known to overestimate both the magnetic moments of the Fe atoms and the Curie temperature \cite{Houlong,ghosh2023unraveling}, which were significantly reduced in terms of dynamical mean-field theory (DMFT) \cite{ghosh2023unraveling}.
Despite recent theoretical efforts, the origin of the relatively large difference between experimentally measured and theoretically predicted $T_\mathrm{C}$ remained unresolved. 

In this work, we systematically investigate the effect of biaxial strain on the Curie temperature and the magnetic phases of a monolayer of $\mathrm{Fe_{3}GeTe_{2}}$. This approach is motivated by the fact that the lattice constant is overestimated in DFT calculations \cite{ghosh2023unraveling}, therefore, it is tempting to investigate the change of the magnetic properties of the system at least in a narrow range of the lattice constants around the theoretical equilibrium value. This may also predict how the critical temperature changes under compressive or tensile strains under experimental conditions. Our theoretical approach is based on combined DFT calculations and spin-model simulations to address the following issues: $(i)$ the variation of magnetic exchange interactions under biaxial strain,  
$(ii)$ the role of competing exchange interactions in determining the magnetic ground state and $(iii)$ the variation of transition temperatures, as well as of the magnetic orderings at finite temperatures. 

The paper is organized as follows: in Section \ref{sec:methods}, we describe the methods and computational details used in this work. In Section \ref{sec:results}, we present the evolution of the magnetic interactions, magnetic ground states, and transition temperatures under biaxial strain.
In addition, we introduce a simple toy model that captures the essential magnetic exchange interactions and provides a transparent framework for analyzing the strain-dependent magnetic ground state. We also highlight the role of Dzyaloshinskii-Moriya interactions (DMI) in stabilizing non-collinear magnetic orderings found for compressive strains.
Finally, in Section \ref{sec:level4}, we summarize our findings and draw conclusions.


\section{\label{sec:methods}Methods}

\subsection{Self-consistent calculations}
DFT calculations were performed using the {\sc siesta} package \cite{siesta}. The exchange-correlation energy was treated within the generalized gradient approximation (GGA) using the Perdew-Burke-Ernzerhof (PBE) functional \cite{PBE}. A $64\times 64\times 1$ Monkhorst-Pack k-point grid for reciprocal-space integrals and a mesh cut-off of $2500$~Ry for real-space integrals guaranteed accurate self-consistent convergence. A double-$\zeta$ polarized basis set was used, and Troullier-Martins norm-conserving pseudopotentials \cite{Troullier} have been employed to describe the interactions between the valence electrons and the atomic cores. The relative error of the density matrix and the error of the Hamiltonian matrix elements have been set to $10^{-6}$ and $10^{-5}$~eV, respectively. The calculations were performed in the ferromagnetic state and spin-orbit coupling \cite{Cuadrado_2012} was taken into account for all systems considered in this work. The size of the unit cell in the out-of-plane direction was set to $20$~\AA{}. For each strained structure, the lattice vectors were kept fixed, and the atomic coordinates were relaxed until the maximum force on each atom was less than $10^{-3}$~eV/\AA{}.

\subsection{Classical spin model}
  In order to study the magnetic structure 
  of the Fe$_3$GeTe$_2$ monolayers, we considered the generalized Heisenberg model,  
\begin{equation}
\begin{split}
H(\textbf{e}_{i}) &= \frac{1}{2}\sum_{i\neq j}J_{ij} \, \textbf{e}_{i}\cdot \textbf{e}_{j} + \frac{1}{2}\sum_{i\neq j}\textbf{e}_{i}\mathcal{J}^{S}_{ij}\textbf{e}_{j} \\
  & + \frac{1}{2}\sum_{i\neq j}\textbf{D}_{ij}\cdot (\textbf{e}_{i}\times \textbf{e}_{j}) + \sum_{i}\textbf{e}_{i}\mathcal{K}_i\textbf{e}_{i}  \, ,\label{Hamiltonian}
\end{split}
\end{equation}

\noindent where $\textbf{e}_{i}$ and $\textbf{e}_{j}$ are classical spin vectors of unit length on sites $i$ and $j$, $J_{ij}$, $\mathcal{J}^{S}_{ij}$, $\textbf{D}_{ij}$ and $\mathcal{K}_{i}$ are isotropic Heisenberg interactions, traceless symmetric exchange matrices, DM 
vectors and on-site anisotropy matrices, respectively. 

The exchange and anisotropy interaction parameters were extracted based on the magnetic force theorem \cite{LIECHTENSTEIN198765,LUdvardi}, as extended to non-orthogonal basis sets \cite{Laci-2019,gabriel} and implemented in the {\sc grogu} package \cite{grogupy}, with the electronic Hamiltonian from the {\sc siesta} calculations serving as input.   
Brillouin zone integrations were performed using a $350 \times 350 \times 1$ Monkhorst-Pack $k$-grid and 50 points were used along a semicircular contour for the energy integrations, ensuring convergence of the exchange and on-site anisotropy parameters to a relative accuracy of $10^{-3}$. The interactions were evaluated for all pairs within a cutoff distance of $5a$, where 
$a$ is the 2D lattice constant. When evaluating the exchange interactions, the Green's functions were projected onto the Fe 3$d$ orbitals to avoid undesired contributions from nonmagnetic orbitals \cite{gabriel}.  

\subsection{Atomistic spin-dynamics simulations}
\label{sec:SD}
To study finite-temperature magnetism, we performed atomistic spin dynamics (ASD) simulations based on the stochastic Landau-Lifshitz-Gilbert (LLG) equation  \cite{Landau:437299,Nowak2007,Rozsa_2014},
\begin{equation}
\frac{\partial \textbf{e}_i}{\partial t}
=
-\frac{\gamma}{1+\alpha^{2}} \, \textbf{e}_i \times \mathbf{B}_i^{\mathrm{eff}}
-
\frac{\alpha}{1+\alpha^{2}} \, \gamma \, 
\textbf{e}_i \times
\left(
\textbf{e}_i \times \mathbf{B}_i^{\mathrm{eff}}
\right) \, ,
\end{equation}
where
\begin{equation}
    \textbf{B}^\mathrm{eff}_{i} = -\frac{1}{M_{i}}\frac{\partial H}{\partial \textbf{e}_{i}} + \boldsymbol{\xi}_i 
\end{equation}
is the effective magnetic field, including a Gaussian white
noise $\boldsymbol{\xi}_i$, $M_{i}$ is the magnetic moment, $\alpha$ is the damping parameter and $\gamma$ is the gyromagnetic factor.
The simulations were performed on a lattice containing $63\times63$ unit cells with periodic boundary conditions. This choice was necessary in order to treat a $\sqrt{3} \times \sqrt{3}$ magnetic unit cell on the hexagonal lattice accurately. At each temperature, the stochastic LLG equations were integrated using a time step of $0.242$~fs for a total of $10^6$ integration steps. The system was thermalized for $5 \times 10^5$ steps, while thermodynamic quantities were averaged over the subsequent $5 \times 10^5$ steps. 

The square root of the averaged squared magnetization, 
\begin{equation}
    \sqrt{\left\langle M^{2}\right\rangle} = \frac{1}{N}\sqrt{\left\langle \sum_{\nu=x,y,z}\left\vert \sum_{i=1}^N e_{i}^{\nu}\right\vert ^{2}\right\rangle}\label{eq:orderparameter}
\end{equation}
is used as an order parameter to trace the magnetic phase transitions, since it is insensitive to collective rotations of the magnetization which commonly occur in 2D systems at higher temperatures~\cite{Jenkins2022}. 
To characterize the magnetic order in the different Fe sublattices, we calculate and present sublattice-resolved order parameters.  

The ground state of the system was determined by zero-temperature LLG spin dynamics simulations \cite{Laszloffy2019} and compared with those obtained from the conjugate gradient method developed in Ref.~\cite{Nagyfalusi_2022}. In this method, the magnetic ground state of the spin model was determined by minimizing the energy based on Eq.~(\ref{Hamiltonian}) with respect to the spin orientations $\{\textbf{e}_{i}\}$, subject to the constraint $\vert \textbf{e}_{i}\vert=1$. The local torque acting on each spin is given by

\begin{equation}
    \textbf{T}_{i}= \textbf{e}_{i} \times \textbf{B}_{i} \, ,
\end{equation}

\noindent where $\textbf{B}_{i}=-\frac{\partial H}{\partial \textbf{e}_{i}}$ determines the direction along which the minimization process is continued. The search directions were updated iteratively according to the Polak-Ribière scheme \cite{polak1969note} ensuring accelerated convergence. The iterations are repeated until the total torque $T=\sqrt{\sum_{i} \textbf{T}_{i}^{2}}$ converged below $10^{-4}$~mRy. For each system, the minimization was started from random initial configurations. For the considered systems, we found that both spin-dynamics simulations and the conjugate gradient method ended in the same spin configuration, 
which is identified as the magnetic ground state.



\section{\label{sec:results}Results}

\subsection{\label{sec:Res-FP} Magnetic moments and spin model parameters}

First, we determined the geometric and electronic structure 
 assuming a ferromagnetic configuration in the DFT calculations. We found the equilibrium lattice constant of the monolayer to be $a = 4.03$~\AA{}, which slightly overestimates the experimental value of $a = 3.99$~\AA{} \cite{roemer2020robust}. It also agrees well with previous DFT calculations, although the theoretical lattice constants scatter in the range of $a = 3.99 - 4.05$~\AA{} depending on the treatment of electron correlations \cite{Houlong}. The calculated magnetic moments for Fe1(Fe2) and Fe3 sublattices are 2.68~$\mathrm{\mu_{B}}$ and 1.51~$\mathrm{\mu_{B}}$, respectively, which also satisfactorily compare with those reported by calculations using GGA for the exchange-correlation potential (2.49~$\mathrm{\mu_{B}}$ and 1.50~$\mathrm{\mu_{B}}$)  \cite{ghosh2023unraveling}. 
These calculations overestimate the experimentally determined \cite{May2016} value for the Fe1(Fe2) magnetic moment, 2.18~$\mathrm{\mu_{B}}$,  while they nicely reproduce the Fe3 moment, 1.54~$\mathrm{\mu_{B}}$. Similar to other DFT studies \cite{Houlong,ghosh2023unraveling}, we confirmed that the GGA+U method leads to largely increased magnetic moments, e.g., to 3.38~$\mathrm{\mu_{B}}$ and 2.71~$\mathrm{\mu_{B}}$ for $U=4$~eV; therefore, we excluded this approach from our further studies.   

\begin{figure}[htb]
\centering
\includegraphics[width=\linewidth]{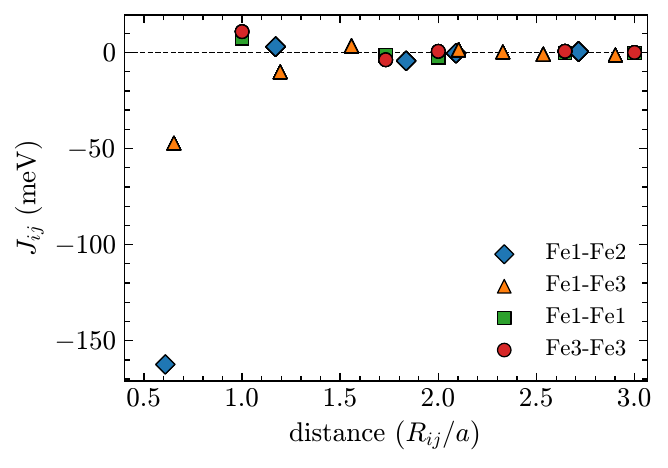}
\vskip -10pt
\caption{Isotropic exchange interactions as a function of distance between the Fe atoms calculated for the equilibrium lattice constant of $a=4.03$~\AA{}. The interactions within or between the inequivalent Fe sublattices are marked by different symbols and colors as shown in the legend. }
\label{fig:jiso_vs_dist}
\end{figure}

Next, we calculated the parameters entering the spin model in Eq.~\eqref{Hamiltonian} by using the {\sc grogu} code \cite{grogupy}. In Fig. \ref{fig:jiso_vs_dist} we present the isotropic exchange interactions as a function of the distance between the Fe atoms. The most important first four interactions are visualized in Fig.~\ref{fig:FGT}(b). The nearest-neighbor (NN) Fe1-Fe2 interaction is strongly ferromagnetic, keeping the spins in these two layers tightly coupled. The second-NN Fe1-Fe3 coupling is also ferromagnetic, while the third NN-Fe1-Fe1 and Fe3-Fe3 interactions are antiferromagnetic and much lower in magnitude. Apparently, the interactions decay very fast and beyond $R_{ij} > 2a$ they can be practically neglected. Taking into account a factor of $-\frac{1}{2}$ in the interaction part of the corresponding spin model, the calculated values of $J_{ij}$ are found to be consistent with those reported in Ref.~\cite{ghosh2023unraveling} within the GGA framework.

\begin{figure}[htb]
\phantomsection\hypertarget{fig:MS_J_D}{}
\includegraphics[width=\linewidth]{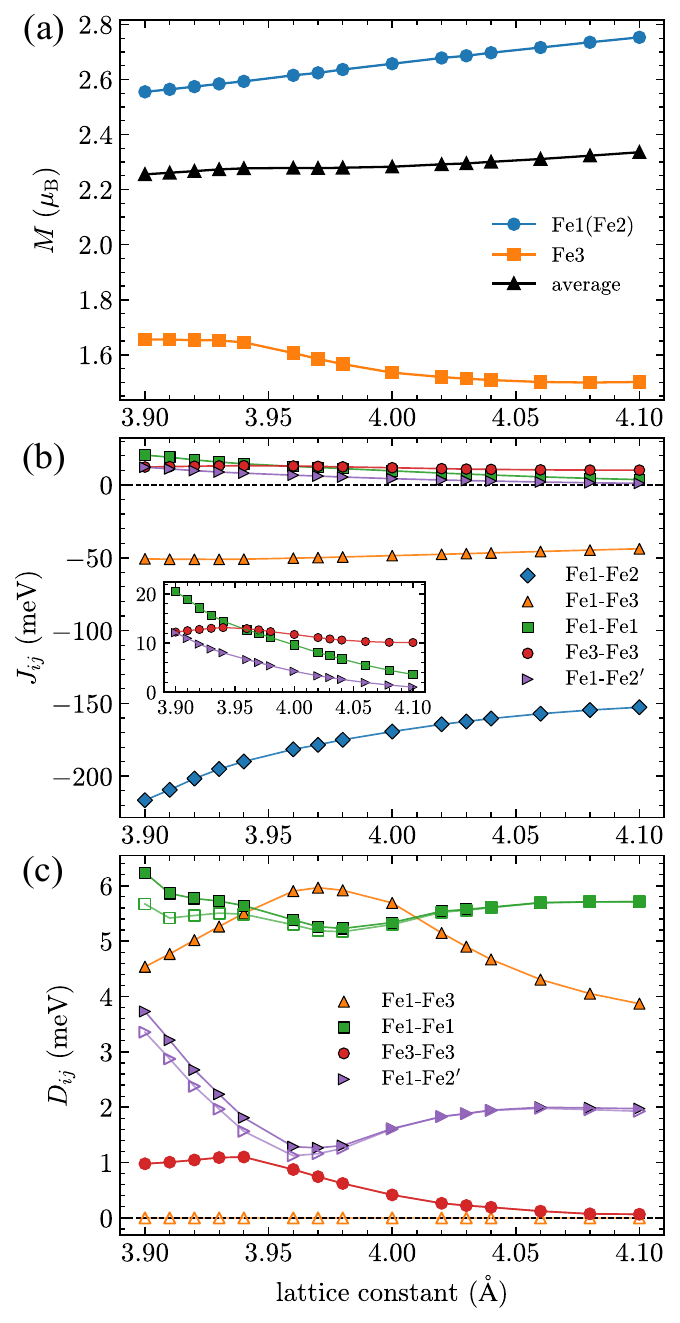}
\vskip -10pt
\caption{(a) Calculated Fe magnetic moments, (b) isotropic exchange interaction and (c) 
DMIs up to fifth-nearest neighbors as a function of the lattice constant $a$. In panel (b), the inset shows an enlarged view of the three closest pairs with AFM couplings. In panel (c), the filled and open symbols stand for the magnitudes of the DM vectors and of their $z$ components, respectively. The NN Fe1-Fe2 DM vector is identically zero by symmetry, see text. }
\label{fig:ms_jiso_dm}
\end{figure}

\begin{figure*}[htb]
    \includegraphics[width=\linewidth]{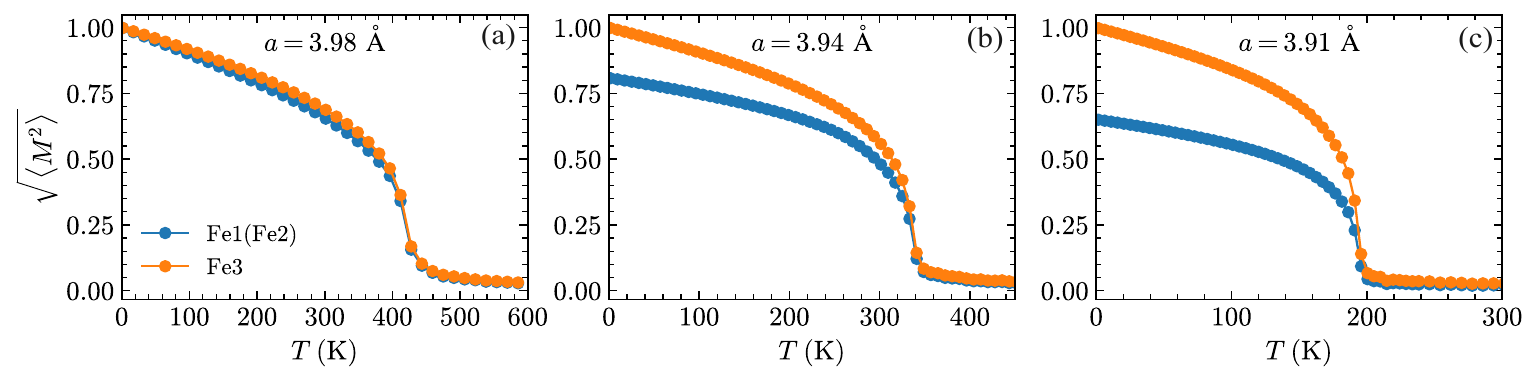}
    \vskip -12pt
    \caption{Order parameters, Eq.~\eqref{eq:orderparameter}, for the two inequivalent Fe sublattices as a function of temperature for three different lattice constants. 
    }
\label{fig:Mnorm}
\end{figure*}

We repeated the calculations by varying the lattice constant from 3.9~\AA{} to 4.1~\AA{}. 
In the DFT calculations, the lengths of the Bravais lattice vectors 
 of the hexagonal lattice were fixed, while the atomic positions were allowed to relax until self-consistency was achieved.
 In all cases, structural relaxation preserved the $D_{3h}$ point-group symmetry of the monolayer.
The calculated magnetic moments for the two inequivalent Fe atoms are shown in Fig.~\ref{fig:ms_jiso_dm}(a). 
The Fe1(Fe2) magnetic moment changes almost linearly with the lattice constant, with a slope of about $1~\mathrm{\mu_{B}}/\mathrm{\AA}$, i.e., decreasing with decreasing lattice constant. In contrast, the magnetic moment of Fe3 increases with decreasing lattice constant,  exhibiting a change of about 0.2~$\mathrm{\mu_{B}}$ in the whole range of lattice constants considered.  
Overall, we observe a very moderate increase of the average Fe moment as the lattice constant increases, which is consistent with the result of Ref.~\cite{Houlong} using the DFT-LDA approach.

The first five NN isotropic exchange couplings are presented in Fig.~\ref{fig:ms_jiso_dm}{(b)} as a function of the lattice constant. 
We find 
NN ferromagnetic interactions between the Fe1-Fe2 and Fe1-Fe3 sublattices, together with antiferromagnetic interactions between Fe-Fe pairs separated by a distance $a$ in the basal $xy$ plane. For this reason, in addition to the NN Fe1-Fe2 pair, we also present results for the Fe1-Fe2$'$ pair, which has an in-plane separation $a$ (not shown in Fig.~\ref{fig:FGT}).
As can be inferred from Fig.~\ref{fig:ms_jiso_dm}(b), all the considered isotropic exchange couplings show a smooth variation against the lattice constant. The strongest FM Fe1-Fe2 increases well beyond 200~meV in magnitude towards $a=3.90$~\AA{}, while the strength of the FM Fe1-Fe3 coupling remains around 50~meV. 
Although the AFM couplings remain small 
compared to the leading FM couplings, in particular, the Fe1-Fe1 and Fe1-Fe2$'$ couplings are significantly strengthened with decreasing lattice constant. The inset of Fig.~\ref{fig:ms_jiso_dm}(b) shows that the magnitude of the Fe1-Fe1 interaction increases from approximately 3~meV to 21~meV, while that of the Fe1-Fe2$'$ interaction increases from about 0.5~meV to 12~meV as the lattice constant decreases from 4.1~\AA{} to 3.9~\AA{}. As shown below, this pronounced strengthening of the AFM interactions plays a key role in driving the system toward a non-collinear magnetic phase and reducing the Curie temperature under compressive strain.
The NN Fe3-Fe3 interaction being in the range of 10~meV -- 14~meV is less sensitive to the strain and it doesn't play a significant role in the formation of the magnetic structure versus strain as the spins in the Fe3 sublattice remain ferromagnetically aligned for all lattice constants.


In  Fig.~\ref{fig:ms_jiso_dm}(c), the magnitudes of the DM vectors and the $z$ components of the DM vectors are shown for the considered Fe-Fe pairs. The DM vector for the NN Fe1-Fe2 pair vanishes due to a mirror plane containing the bond and a $C_{3}$ symmetry axis around the bond. The presence or absence of the $z$ components are consistent with Moriya’s rules \cite{Moriya}. For nearest-neighbor Fe1-Fe3 pairs, the DMI vectors have no $z$ component due to the mirror plane containing the bond. For the NN Fe3-Fe3 pairs, the DM vectors are purely out-of-plane due to a $xy$ mirror plane containing the bond. In contrast, the DM vectors associated with the Fe1-Fe1 and Fe1-Fe2$'$ pairs are predominantly out of plane, with a small but symmetry-allowed in-plane component.  
Notably, the large magnitudes of the Fe1-Fe1 and Fe1-Fe2$'$ DM vectors, particularly of their $z$-components, are essential for stabilizing the emerging non-collinear spin structures, as we will show later.   

Based on the spin-model Eq.~\eqref{Hamiltonian}, we calculated the magnetic anisotropy energy (MAE) per unit cell, 
\begin{equation}
E_{x}-E_{z} = \sum_{i \in \mathrm{unit \ cell}} \left( \frac{1}{2}\sum_{j} (J^{xx}_{ij} - J^{zz}_{ij}) + (K^{xx}_{i}-K^{zz}_{i}) \right) \, .   
\end{equation}
We found that the system exhibits robust out-of-plane magnetic anisotropy, with the MAE ranging from 3.5 to 6~meV per unit cell over the range of lattice constants considered. These values are in good agreement with the MAE of approximately $0.6-4.5$~meV per unit cell reported in Ref.~\cite{Houlong} for strains between $-4$~\% and 4~\%, as obtained from DFT-LDA calculations.

\subsection{\label{sec:Res-SD} Magnetism of strained Fe$_3$GeTe$_2$ monolayers}

Using the calculated spin-model parameters we performed ASD simulations for all lattice constants. In Fig.~\ref{fig:Mnorm} we plotted the order parameters defined in Eq. \eqref{eq:orderparameter} for the Fe1(Fe2) and the Fe3 sublattices as a function of temperature for three selected lattice constants. For $a=3.98$~\AA{}, the order parameters approach 
$\sqrt{\left\langle M^{2}\right\rangle} =1$ for $T=0$~K, indicating that the ground state is ferromagnetic in all Fe sublattices. This is the case also for any lattice constant larger than 3.98~\AA{}. In the case of $a=3.94$~\AA{}, see Fig.~\ref{fig:Mnorm}(b), the Fe1(Fe2) order parameter approaches a value 
$\sqrt{\left\langle M^{2}\right\rangle} <1$ at $T=0$~K, indicating that in the ground state the ordering in the Fe1(Fe2) sublattices is not ferromagnetic. For even smaller lattice parameters, see Fig.~\ref{fig:Mnorm}(c) for $a=3.91$~\AA{}, the value of the Fe1(Fe2) order parameter further decreases. It is important to note that, independent of the strain, the Fe3 sublattice remains ferromagnetic and the order parameters for the two inequivalent Fe sublattices vanish at the same temperature $T_\mathrm{C}$. Moreover, $T_\mathrm{C}$ rapidly decreases with decreasing lattice constant: $T_\mathrm{C}\simeq420$~K for $a=3.98$~\AA{}, $T_\mathrm{C}\simeq340$~K for $a=3.94$~\AA{} and $T_\mathrm{C}\simeq190$~K for $a=3.91$~\AA{}. 

\medskip
\begin{figure}[htb]
    \phantomsection\hypertarget{fig:grdstate}{}
    \centering
    \includegraphics[width=0.9\linewidth]{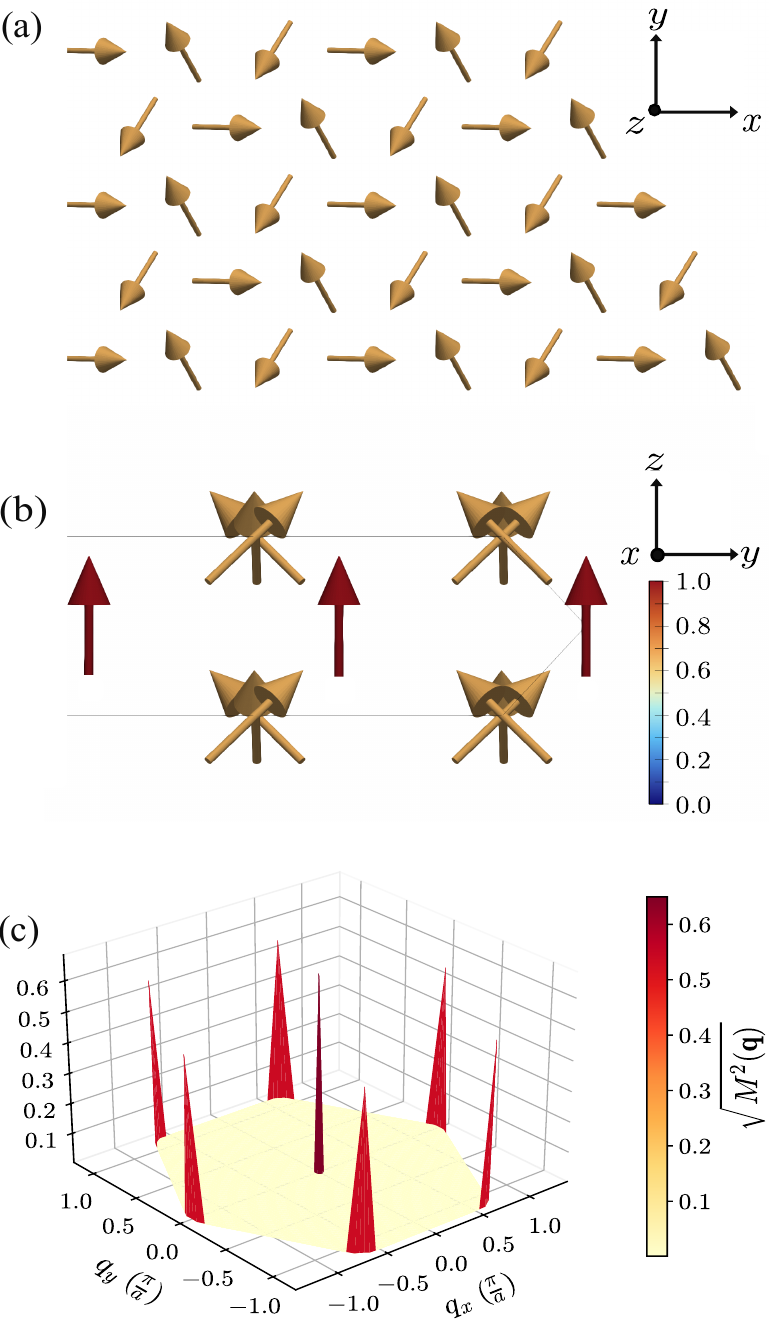}
    \caption{(a) Top and (b) side view of the ground state for lattice constant $a = 3.91$~\AA{}. 
Panel (a) shows only the spin vectors in the Fe1 layer, whereas (b) includes the three Fe layers. 
The color bar, common to both (a) and (b) panels, stands for the normalized $z$ component of the spins. (c) Distribution of the 
static structure factor in the Brillouin zone, Eq.~\eqref{eq:orderparameter-q}, for the Fe1(Fe2) sublattice at zero temperature.}
\label{fig:grdstate}
\end{figure}

To understand the variation of the order parameter at zero temperature against the strain, we simulated the magnetic ground state of the Fe$_3$GeTe$_2$ monolayer as outlined in Section \ref{sec:SD}. For lattice constants $a \ge 3.98$~\AA{}, we found a ferromagnetic ground state in which all spins are aligned along the $z$ direction. This is an obvious consequence of strong FM exchange interactions, see Fig.~\ref{fig:ms_jiso_dm}(b), and large perpendicular magnetic anisotropy. For lattice constants $a \le 3.97$~\AA{}, the magnetic ground state of the system changes markedly: the spins in the Fe3 sublattice remain aligned along the $z$ axis, while the Fe1 and Fe2 sublattices develop a conical spin-spiral (CSS) configuration with three spins forming the magnetic unit cell. As an example, the ground-state spin structure for $a = 3.91$~\AA{} is depicted in Fig.~\ref{fig:grdstate}(a) and (b). 
Remarkably, the spin configurations in the Fe1 and Fe2 sublattices are perfectly aligned, i.e., the Fe spins at sites connected by a bond along the $z$ axis are parallel to each other, owing to the extremely strong NN FM coupling between them. 

The ground-state spin configuration shown in Fig.~\ref{fig:grdstate}(a) and (b) can be rotated by arbitrary angles around the $z$ axis without any cost of energy, which means that the ground state of the bilinear spin-Hamiltonian Eq.~\eqref{Hamiltonian} is continuously degenerate. Although this degeneracy can be lifted by including higher-order on-site anisotropy terms, the order parameter introduced in Eq.~\eqref{eq:orderparameter} provides an efficient means of investigating the finite-temperature magnetism of the system even in case of continuous degeneracy. 
We also note that the chirality of the CSS is defined as the sign of the $z$ component of the chirality vector $\mathbf{e}_1 \times \mathbf{e}_2 + \mathbf{e}_2 \times \mathbf{e}_3 + \mathbf{e}_3 \times \mathbf{e}_1$, where $\mathbf{e}_i$ $(i=1,2,3)$ are the spin vectors in a magnetic unit cell, ordered in a clockwise sequence. As can be seen in 
Fig.~\ref{fig:grdstate}(a), the chirality of the ground state CSS is $-1$, fixed by the positive sign of the $z$ component of the NN Fe1-Fe1 DM vectors.

In order to characterize the non-collinear magnetic order in the Fe1(Fe2) sublattices quantitatively, we calculated 
the square root of the static structure factor of the spin configuration, defined as \cite{Rozsa2015}
\begin{equation}
\sqrt{M^{2}(\textbf{q})} = \frac{1}{N}\sqrt{\left\langle \sum_{\nu=x,y,z}\left| \sum_{j}e^{i\textbf{q}\textbf{R}_{j}} e_{j}^{\nu}\right|^{2}\right\rangle} \, .\label{eq:orderparameter-q}
\end{equation}
Evidently, the order parameter introduced in Eq.~\eqref{eq:orderparameter} is equivalent to $\sqrt{M^{2}(\textbf{q}=0)}$, which measures the degree of ferromagnetic order in the system.
The results are shown in Fig.~\ref{fig:grdstate}(c) for $a = 3.91$~\AA{}, where distinct peaks can be found at the center of the Brillouin zone ($\Gamma$ point) and at the six corners of the Brillouin zone ($K$ points). Clearly, the $\Gamma$ peak arises from the $z$ component of the spin vectors, while the in-plane projections of the spin vectors contribute to the peaks at the $K$ points. The ratio of the heights of the $K$ and $\Gamma$ peaks is related to the tilting angle of the conical spin spiral.   
Note that the 
structure factor for the Fe3 sublattice exhibits a single peak at the $\Gamma$ point for all lattice constants, confirming the FM order in this Fe sublattice. 

It is well known that antiferromagnetic nearest-neighbor interactions on a triangular lattice constitute a prototypical example of geometrical frustration, resulting in a 120$^\circ$ N\'eel-ordered state. As the lattice constant decreases, the AFM Fe1-Fe1 (Fe2-Fe2) interactions increasingly compete with the FM Fe1-Fe3 interactions and the strong perpendicular magnetic anisotropy, which together favor an out-of-plane ferromagnetic state. Our results indicate that beyond a certain strength of the AFM interaction, this competition leads to the formation of a conical spin-spiral state.        

\begin{figure*}[htb]
    \includegraphics[width=\linewidth]{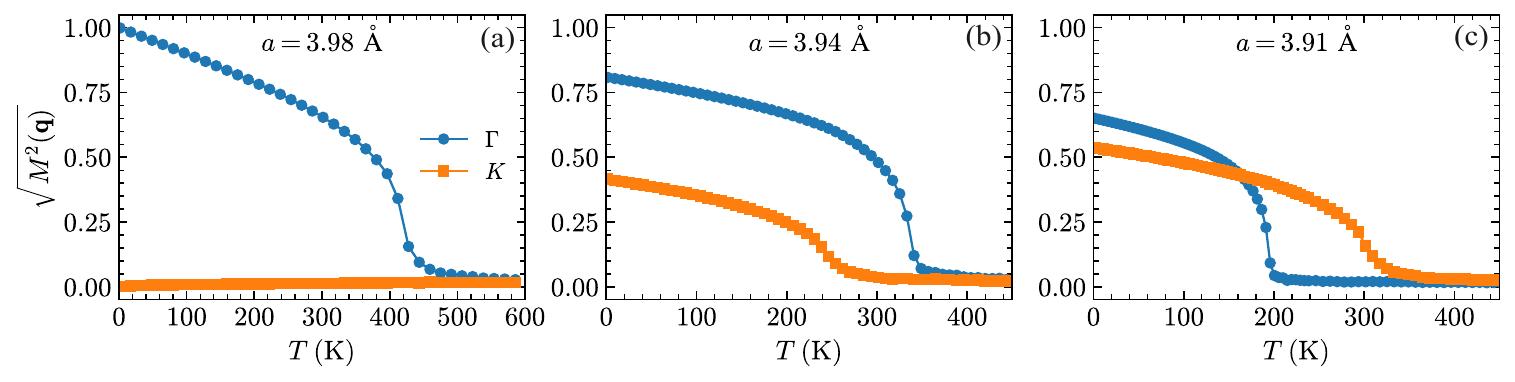}
    \vskip -10pt
    \caption{
    Static structure factors at the $\Gamma$ and $K$ points of the Brillouin zone, see Eq.~\eqref{eq:orderparameter-q}, for the Fe1(Fe2) sublattice as a function of temperature for three different lattice constants. 
    $\sqrt{M^{2}(\Gamma)}$ (blue circles) and $\sqrt{M^{2}(K)}$ (orange squares) represent the out-of-plane and the in-plane components of the magnetization, respectively. 
    }
\label{fig:M_Gamma_K}
\end{figure*}

To obtain a simple yet quantitative understanding of this phase transition, we investigate a minimal spin model for Fe$_3$GeTe$_2$ in Appendix \ref{sec:appendix}, including the first three nearest-neighbor isotropic interactions with parameters, $J_1 = J_{\mathrm{Fe1}-\mathrm{Fe2}} < 0$, $J_2 = J_{\mathrm{Fe1}-\mathrm{Fe3}} < 0$, $J_3 = J_{\mathrm{Fe1}-\mathrm{Fe1}} >0$ and a uniaxial anisotropy parameter, $K<0$, for all the three Fe sublattices. This model indeed predicts the stabilization of the conical spiral ground state over the ferromagnetic state when the following condition is satisfied: 
\begin{equation}
J_{3} > -\frac{1}{3}J_{2}-\frac{2}{9}K \, .
\end{equation}
Moreover, the opening angle of the conical spin spiral is given by
\begin{equation}
\theta_\mathrm{CSS} = \arccos \left( - \frac{J_2}{3 J_3 + \frac{2}{3} K} \right) \, ,
\end{equation}
implying that it increases with increasing in-plane AFM coupling $J_3$. 

In Fig.~\ref{fig:M_Gamma_K}, the temperature dependence of the 
structure factors for the Fe1(Fe2) sublattices at the $\Gamma$ and $K$ points are shown for $a= 3.98$~\AA{}, $3.94$~\AA{} and $3.91$~\AA{}. For $a= 3.98$~\AA{}, $\sqrt{ M^{2}(K)}$ is zero within statistical fluctuations, implying that N\'eel order is not present in the system. This is the case for all lattice constants, $a \ge 3.98$~\AA{}, for which the magnetic ground state is ferromagnetic. For $a= 3.94$~\AA{}, $\sqrt{ M^{2}(K)}$ is finite at $T=0$~K, decreases with increasing temperature and vanishes at about 240~K, which we 
identify as the N\'eel temperature, $T_\mathrm{N}$. In the temperature range $T_\mathrm{N} < T < T_\mathrm{C}$, $\sqrt{ M^{2}(\Gamma)}$ is finite, i.e., the system is in the ferromagnetic phase.  
The case of $a= 3.91$~\AA{} exhibits a different  situation, as the 
structure factor $\sqrt{ M^{2}(K)}$ vanishes at a higher 
temperature than $\sqrt{ M^{2}(\Gamma)}$, i.e., $T_\mathrm{N} > T_\mathrm{C}$. This means that for temperatures, $T_\mathrm{C} < T < T_\mathrm{N}$ the system is in a planar N\'eel-ordered phase. 

\begin{figure}[htb]
\centering
\includegraphics[width=\columnwidth]{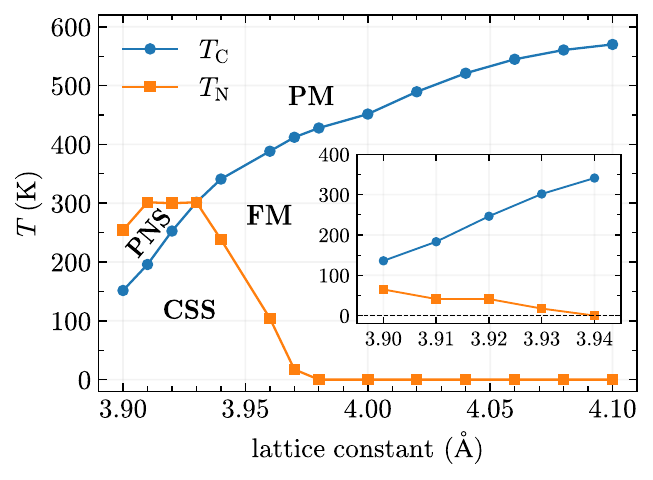}
\vskip -10pt
\caption{Transition temperatures $T_\mathrm{C}$ and $T_\mathrm{N}$ as a function of the lattice constant obtained from atomistic spin-dynamics simulations. The acronyms denote the different magnetic phases: FM -- ferromagnetic, CSS -- conical spin spiral, PNS -- planar N\'eel state, PM -- paramagnetic. 
The inset shows $T_\mathrm{C}$ and $T_\mathrm{N}$ excluding DMI from the spin model for $a = 3.90-3.94$~\AA{}. }
\label{fig:Tc_Tn}
\end{figure}

For all lattice constants considered in our study, the simulated N\'eel and Curie temperatures are shown in Fig.~\ref{fig:Tc_Tn}. 
In the range $3.90 \le a \le 3.92$ \AA{}, increasing the temperature drives the Fe1(Fe2) sublattices first from a conical spin-spiral phase to a planar Néel phase and subsequently to a paramagnetic state.  For $a=3.93$~\AA{}, we observed a single transition from a conical spin-spiral to a paramagnetic state. For $3.94 \le a \le 3.97$ \AA{}, the Fe1(Fe2) sublattices exhibit a sequence of transitions from a conical spin-spiral to a ferromagnetic and then to a paramagnetic phase. For lattice constants $a \geq 3.98$~\AA{}, only a single transition from a ferromagnetic to a paramagnetic phase occurs.

As it is clear from Fig.~\ref{fig:ms_jiso_dm}(c), the magnitude of the nearest-neighbor DM 
vectors in the Fe1(Fe2) sublattices is about $5-6$~meV, and they are oriented almost normal to the plane. Therefore, the DMI is expected to make a significant contribution to both the energy of the conical spin-spiral state and the thermal stability of the corresponding phase. To assess this effect, we performed LLG simulations with the DMI terms removed from the spin Hamiltonian, Eq.~\eqref{Hamiltonian}. The results for the Curie and the N\'eel temperatures are shown in the inset of Fig.~\ref{fig:Tc_Tn} for $a \in [3.90~\mathrm{\AA{}},3.94~\mathrm{\AA{}}]$. 
In the absence of DMI, the Néel temperature is significantly reduced and becomes smaller than the Curie temperature. This means that by increasing the temperature, the Fe1 and Fe2 sublattices undergo a transition from a conical spin-spiral to a ferromagnetic and subsequently to a paramagnetic phase, i.e., the planar N\'eel phase is no longer stabilized. The Curie temperature remains unaffected by the removal of DMI, except at $a=3.90$~\AA{}, where a reduction of approximately 10~K is observed.

Finally, we relate our theoretical results to the available experimental findings. Dang {\em et al.}~\cite{MossbauerSpectro} performed synchrotron Mössbauer spectroscopy measurements on single-crystalline bulk Fe$_3$GeTe$_2$, and observed a pronounced suppression of both the hyperfine magnetic fields and the Curie temperature with increasing pressure. Furthermore, a paramagnetic ground state emerged above a critical pressure of approximately 15~GPa. The rapid decrease of the Curie temperature under compressive strain obtained from our LLG simulations is consistent with these observations.

\begin{figure}[htb]
\centering
\includegraphics[width=\columnwidth]{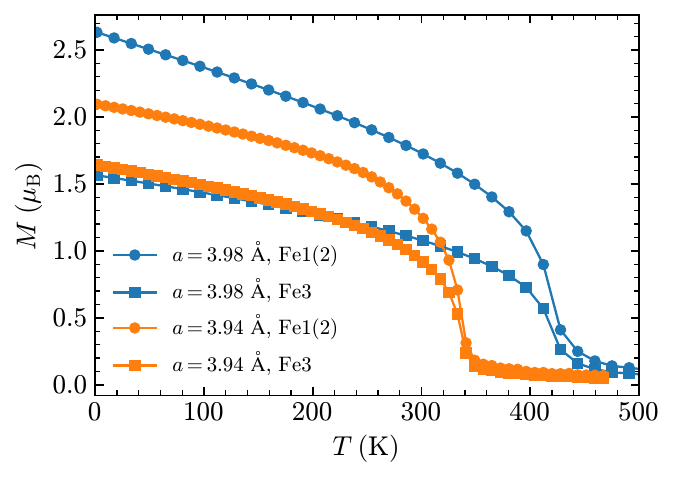}
\vskip -10 pt
\caption{Temperature dependence of the magnetic moments in the inequivalent sublattices, defined as the order parameter 
multiplied by the magnetic moment from the DFT calculation, for $a = 3.94$~\AA{} and 3.98~\AA{}.}
\label{fig:M_vs_T}
\end{figure}

Remarkably, the hyperfine field at the Fe3 site (denoted Fe2 in Ref.~\cite{MossbauerSpectro}) decreases only gradually with increasing pressure, whereas the hyperfine field at the Fe1 site is suppressed much more rapidly at moderate pressures. Fig.~\ref{fig:M_vs_T} shows the temperature dependence of the Fe magnetic moments, calculated as the product of the order parameter and the magnetic moment obtained from the DFT calculation, for two different lattice constants. While the magnetic moment at the Fe3 sites at zero temperature remains essentially unchanged, the moment at the Fe1(Fe2) sites is significantly reduced for the smaller lattice constant. The temperature dependence of the hyperfine fields at the two inequivalent Fe sites was also reported in Ref.~\cite{MossbauerSpectro} for two different pressures, exhibiting a remarkable similarity to Fig.~\ref{fig:M_vs_T}.
Therefore, the possibility cannot be excluded that the pressure-induced reduction of the hyperfine field at the Fe1(2) site is, at least in part, associated with the transition from the ferromagnetic phase to the conical spin-spiral phase.

\section{\label{sec:level4}
Conclusions}

In this work, the magnetic properties of $\mathrm{Fe_{3}GeTe_{2}}$ monolayer were systematically investigated as a function of lattice constant. The extracted magnetic exchange parameters revealed a delicate balance between FM and AFM interactions, with AFM coupling becoming increasingly dominant under lattice compression. 
Atomistic spin-dynamics simulations show that the Curie temperature decreases with decreasing lattice constant, accompanied by an enhanced disparity between the magnetization of the two inequivalent sublattices. While the Fe3 sublattice remains ferromagnetic with out-of-plane magnetization, the Fe1(Fe2) sublattices progressively deviate from collinearity.
The magnetic ground state undergoes a transition from a collinear ferromagnetic configuration to a noncollinear conical spiral state in the Fe1(Fe2) sublattices under compressive strain.

We analyzed the temperature dependence of the 
static structure factor and found multiple thermally driven phase transitions including transitions between conical spin-spiral, planar N\'eel-ordered, ferromagnetic, and paramagnetic phases at finite temperatures. 
Furthermore, we investigated the role of the DMI in the ordering temperatures. While $T_\mathrm{C}$ remains largely unaffected when the DMI is turned off, the N\'eel temperature is remarkably reduced at smaller lattice constants, and the planar N\'eel phase is no longer stabilized. 

Overall, these results highlight the intricate interplay between isotropic interactions, DMI, and magnetic anisotropy in determining the magnetic ground state and finite-temperature magnetism in the $\mathrm{Fe_{3}GeTe_{2}}$ monolayer. These findings provide insight into the microscopic mechanisms underlying non-collinear magnetism and may guide the design of materials with tunable magnetic phases under strain.

\section{Acknowledgments}
The work of A. J. B., B. N., L. U., L. R. and L. S. was supported by the Ministry of Culture and Innovation and the National Research, Development and Innovation Office under Grant Nos. K142652, FK142601 and ADVANCED 149745. L. R., B. N. and Z. T. acknowledge funding by the Hungarian Academy of Sciences via J\'{a}nos Bolyai Research Grants (Grant Nos. BO/00178/23/11, BO/00893/25 and BO/00070/24).
This project was also supported by the TRILMAX Horizon Europe consortium (Grant No. 101159646). G. M.-C., A. G.-F. and J. F. have been funded by Ministerio de Ciencia, Innovación y Universidades, Agencia
Estatal de Investigación, Fondo Europeo de Desarrollo Regional via the grant PID2022-137078NB-I00,
and by Agencia SEKUENS (Asturias) under grant UONANO IDE/2024/000678 with the support of FEDER funds.
Most of the calculations in this work were performed using the Komondor HPC facility in Hungary, operated by Digital Government Development and Project Management Ltd.

\section{Data Availability}
The data that support the findings of this study, including the raw outputs of {\sc siesta}, {\sc grogu} and LLG dynamics calculations are openly available at \cite{jyothi_bhasu_2026_20737279}.

\appendix

\section{\label{sec:appendix} 
Toy model of the magnetic ground state of  Fe$_{3}$GeTe$_{2}$}

The primitive lattice vectors of a 
triangular lattice and its reciprocal lattice are given as
\begin{align}
\mathbf{a}_1 &= a(1,0), &
\mathbf{a}_2 &= a\left(-\frac{1}{2},\frac{\sqrt{3}}{2}\right) \, ,
\end{align}
and
\begin{align}
\mathbf{b}_1 &= \frac{2\pi}{a}\left(1,\frac{1}{\sqrt{3}}\right), &
\mathbf{b}_2 &= \frac{2\pi}{a}\left(0,\frac{2}{\sqrt{3}}\right) \, ,
\end{align}
respectively. 
The lattice positions in the Fe1, Fe3 and Fe2 sublattices can be written as:%
\begin{align*}
\mathbf{R}_1 &= \mathbf{R}
= n\mathbf{a}_1 + m\mathbf{a}_2,
\qquad n,m\in\mathbb{Z},
\\[4pt]
\mathbf{R}_3 &= \mathbf{R}
+ \frac{2\mathbf{a}_1+\mathbf{a}_2}{3}
+ \mathbf{c},
\\[4pt]
\mathbf{R}_2 &= \mathbf{R} + 2\mathbf{c},
\end{align*}
where
\begin{equation*}
\mathbf{c}=(0,0,-c) \, .
\end{equation*}
A spin vector in sublattice $\alpha$ and unit cell with lattice vector $\mathbf{R}_\alpha$ will be denoted by
\begin{equation*}
\mathbf{e}_{\alpha,\mathbf{R}}
=
\left(
e_{\alpha,\mathbf{R}}^{x},
e_{\alpha,\mathbf{R}}^{y},
e_{\alpha,\mathbf{R}}^{z}
\right),
\end{equation*}
with the constraint,
\begin{equation*}
\left|\mathbf{e}_{\alpha,\mathbf{R}}\right|=1,
\qquad \alpha=1,2,3 \, .
\end{equation*}

In the spin Hamiltonian we consider the first three nearest-neighbour isotropic interactions and on-site anisotropy:
\begin{equation}
H = H_{12} + H_{31} + H_{32} + H_{11} + H_{22} + H_a \, . \label{Energy}
\end{equation}
with
\begin{align*}
H_{12}
&=
J_1 \sum_{\mathbf{R}}
\mathbf{e}_{1,\mathbf{R}}
\mathbf{e}_{2,\mathbf{R}},
\\[4pt]
H_{31}
&=
J_2 \sum_{\mathbf{R}}
\mathbf{e}_{3,\mathbf{R}}
\Big(
\mathbf{e}_{1,\mathbf{R}}
+
\mathbf{e}_{1,\mathbf{R}+\mathbf{a}_1}
+
\mathbf{e}_{1,\mathbf{R}+\mathbf{a}_1+\mathbf{a}_2}
\Big),
\\[4pt]
H_{32}
&=
J_2 \sum_{\mathbf{R}}
\mathbf{e}_{3,\mathbf{R}}
\Big(
\mathbf{e}_{2,\mathbf{R}}
+
\mathbf{e}_{2,\mathbf{R}+\mathbf{a}_1}
+
\mathbf{e}_{2,\mathbf{R}+\mathbf{a}_1+\mathbf{a}_2}
\Big),
\end{align*}

\begin{align*}
H_{11}
&=
\frac{J_3}{2}
\sum_{\mathbf{R}}
\mathbf{e}_{1,\mathbf{R}}
\Big(
\mathbf{e}_{1,\mathbf{R}+\mathbf{a}_1}
+
\mathbf{e}_{1,\mathbf{R}+\mathbf{a}_2}
+
\mathbf{e}_{1,\mathbf{R}+\mathbf{a}_1+\mathbf{a}_2}
\nonumber\\
&\hspace{2.0cm}
+
\mathbf{e}_{1,\mathbf{R}-\mathbf{a}_1}
+
\mathbf{e}_{1,\mathbf{R}-\mathbf{a}_2}
+
\mathbf{e}_{1,\mathbf{R}-\mathbf{a}_1-\mathbf{a}_2}
\Big),
\\[6pt]
H_{22}
&=
\frac{J_3}{2}
\sum_{\mathbf{R}}
\mathbf{e}_{2,\mathbf{R}}
\Big(
\mathbf{e}_{2,\mathbf{R}+\mathbf{a}_1}
+
\mathbf{e}_{2,\mathbf{R}+\mathbf{a}_2}
+
\mathbf{e}_{2,\mathbf{R}+\mathbf{a}_1+\mathbf{a}_2}
\nonumber\\
&\hspace{2.0cm}
+
\mathbf{e}_{2,\mathbf{R}-\mathbf{a}_1}
+
\mathbf{e}_{2,\mathbf{R}-\mathbf{a}_2}
+
\mathbf{e}_{2,\mathbf{R}-\mathbf{a}_1-\mathbf{a}_2}
\Big),
\end{align*}
and
\begin{equation*}
H_a
=
K
\sum_{\alpha=1}^{3}
\sum_{\mathbf{R}}
\left(
e_{\alpha,\mathbf{R}}^{z}
\right)^2 \, ,
\end{equation*}
\noindent where $J_{1}<0$ is the ferromagnetic first-NN interaction between the Fe1 and
Fe2 sublattices, $J_{2}<0$ is the ferromagnetic second-NN interaction between sublattices Fe3 and Fe1
(Fe2), while $J_{3}>0$ is the third-NN AFM in-plane interaction in the Fe1 and Fe2 sublattices. Note that the terms with interactions between different sublattices contain no double counting, while in $H_{11}$ and $H_{22}$ each
interaction occurs twice, this explains the prefactor of $\frac{1}{2}$. We
consider a uniform out-of-plane on-site anisotropy $K<0$ for all sublattices.

The energy per unit cell of the ferromagnetic state with a
magnetization along the \emph{z} axis can easily be derived as, %
\begin{equation}
E_{\mathrm{FM}}=J_{1}+6J_{2}+6J_{3}+3K \,.
\end{equation}

A conical spin-spiral 
state, in which the spins are rotating around the $z$ axis with an opening angle of $\theta$, is described as
\begin{align*}
e_{1,\mathbf{R}}^{x}
&=
\sin\theta\,
\cos(\mathbf{q}\cdot\mathbf{R}),
\\
e_{1,\mathbf{R}}^{y}
&=
\sin\theta\,
\sin(\mathbf{q}\cdot\mathbf{R}),
\\
e_{1,\mathbf{R}}^{z}
&=
\cos\theta  \, ,
\end{align*}
where $\mathbf{q}=(q_x,q_y)$  is the wave vector of the spin spiral.

Since the FM interaction $J_{1}$ is several times larger in magnitude than
$J_{2}$ and $J_{3}$, we suppose identical conical spin-spiral states 
in the Fe1 and Fe2 sublattices,
and a ferromagnetic state in the Fe3 sublattice. The energy per unit cell in the conical spin-spiral state can then be expressed as
\begin{align}
E_{\mathrm{CSS}}(\theta,\mathbf{q}) & = J_1 + K + 6J_2\cos\theta + (6J_3+2K)\cos^2\theta \nonumber\\ &\quad 
+ 2J_3\sin^2\theta \Big[ \cos(\mathbf{q}\cdot \mathbf{a}_1) + \cos(\mathbf{q}\cdot\mathbf{a}_2)
\nonumber\\
&\quad + \cos\big(\mathbf{q}\cdot(\mathbf{a}_1+\mathbf{a}_2)\big) \Big].
\end{align}
$E_{\mathrm{CSS}}(\theta,\mathbf{q})$ is minimized at the $K$ point of the Brillouin zone, i.e., for $\mathbf{q}_K= \frac{2\mathbf{b}_1-\mathbf{b}_2}{3}$:
\begin{align}
E_{\mathrm{CSS}}(\theta,\mathbf{q}_K) = & \, J_1 -3 J_3 + K + 6J_2\cos\theta \nonumber \\ 
 & + (9J_3+2K)\cos^2\theta \, .
\end{align}

Expanding the energy up to second order in $\theta$,
\begin{align}
E_{\mathrm{CSS}}(\theta,\mathbf{q}_K) = E_{\mathrm{FM}}
- \left(3J_2+9J_3+2K\right) 
\theta^2 \, ,
\end{align}
implies that for
\begin{equation}
J_{2}+3J_{3}+\frac{2}{3}K<0
\end{equation}
the energy of the ferromagnetic state is lower than the energy of the conical spin-spiral state, while
for%
\begin{equation}
J_{2}+3J_{3}+\frac{2}{3}K>0
\end{equation}
the energy of the conical spin-spiral state is lower than the energy of the ferromagnetic state. 

To determine the angle that minimizes the energy of the conical spin-spiral
state with $\textbf{q}_{K}$,
we have to solve the equation
\[
\frac{\textrm{d}E_{\mathrm{CSS}}\left(  \theta,\textbf{q}_{K}\right)
}{\textrm{d}\theta}=-6J_{2}\sin\theta-(9J_{3}+2K)\sin2\theta=0\,
\]
from which we get
\begin{align}
J_{2}+(3J_{3}+\frac{2}{3}K)\cos\theta=0\,.
\end{align}
Since $J_{2} +3J_{3}+\frac{2}{3}K>0$, there exists a tilting angle,
\begin{align}
\theta_{\mathrm{CSS}}  &  = \mathrm{arccos} \left( -\frac{J_{2}}{3J_{3}+\frac{2}{3}K} \right) \,,
\end{align}
for which $E_{\mathrm{CSS}}\left(  \theta,\textbf{q}_{K}\right)$ is minimized.

\bibliography{FGT}

@PREAMBLE{
 "\providecommand{\noopsort}[1]{}" 
 # "\providecommand{\singleletter}[1]{#1}%" 
}

@article{huang2017layer,
  title={{Layer-dependent ferromagnetism in a van der Waals crystal down to the monolayer limit}},
  author={Huang, Bevin and Clark, Genevieve and Navarro-Moratalla, Efren and
            Klein, Dahlia R. and Cheng, Ran and Seyler, Kyle L. and
            Zhong, Ding and Schmidgall, Emma and McGuire, Michael A. and
            Cobden, David H. and Yao, Wang and Xiao, Di and
            Jarillo-Herrero, Pablo and Xu, Xiaodong},
  doi          = {10.1038/nature22391},
  url          = {https://doi.org/10.1038/nature22391},
  journal={Nature},
  volume={546},
  number={7657},
  pages={270--273},
  year={2017},
  publisher={Nature Publishing Group UK London}
}

@article{gong2017discovery,
  title={{Discovery of intrinsic ferromagnetism in two-dimensional van der Waals crystals}},
  author={Gong, Cheng and Li, Lin and Li, Zhenglu and Ji, Huiwen and
            Stern, Alex and Xia, Yang and Cao, Ting and Bao, Wei and
            Wang, Chenzhe and Wang, Yuan and Qiu, Z. Q. and
            Cava, Robert J. and Louie, Steven G. and Xia, Jing and
            Zhang, Xiang},
  doi          = {10.1038/nature22060},
  url          = {https://doi.org/10.1038/nature22060},
  journal={Nature},
  volume={546},
  number={7657},
  pages={265--269},
  year={2017},
  publisher={Nature Publishing Group UK London}
}

@article{yang2022two,
  title={{Two-dimensional materials prospects for non-volatile spintronic memories}},
  author={Yang, Hyunsoo and Valenzuela, Sergio O. and
            Chshiev, Mairbek and Couet, Sebastien and
            Dieny, Bernard and Dlubak, Bruno and
            Fert, Albert and Garello, Kevin and
            Jamet, Matthieu and Jeong, Dae-Eun and
            Lee, Kangho and Lee, Taeyoung and
            Martin, Marie-Blandine and Kar, Gouri Sankar and
            S\'en\'eor, Pierre and Shin, Hyeon-Jin and
            Roche, Stephan},
  doi          = {10.1038/s41586-022-04768-0},
  url          = {https://doi.org/10.1038/s41586-022-04768-0},
  journal={Nature},
  volume={606},
  number={7915},
  pages={663--673},
  year={2022},
  publisher={Nature Publishing Group UK London}
}

@article{fert2024electrical,
  title={{Electrical control of magnetism by electric field and current-induced torques}},
  author={Fert, Albert and Ramesh, Ramamoorthy and
            Garcia, Vincent and Casanova, Felix and
            Bibes, Manuel},
  journal={Reviews of Modern Physics},
  doi          = {10.1103/RevModPhys.96.015005},
  url          = {https://doi.org/10.1103/RevModPhys.96.015005},
  volume={96},
  number={1},
  pages={015005},
  year={2024},
  publisher={APS}
}

@article{mermin1966absence,
  title={{Absence of ferromagnetism or antiferromagnetism in one-or two-dimensional isotropic Heisenberg models}},
  author={Mermin, N David and Wagner, Herbert},
  journal={Physical Review Letters},
  doi          = {10.1103/PhysRevLett.17.1133},
  url          = {https://doi.org/10.1103/PhysRevLett.17.1133},
  volume={17},
  number={22},
  pages={1133},
  year={1966},
  publisher={APS}
}

@article{xu2018interplay,
  title={{Interplay between Kitaev interaction and single ion anisotropy in ferromagnetic $\mathrm{CrI}_3$ and $\mathrm{CrGeTe}_3$ monolayers}},
  author={Xu, Changsong and Feng, Junsheng and Xiang, Hongjun and Bellaiche, Laurent},
  journal={npj Computational Materials},
  doi          = {10.1038/s41524-018-0115-6},
  url          = {https://doi.org/10.1038/s41524-018-0115-6},
  volume={4},
  number={1},
  pages={57},
  year={2018},
  publisher={Nature Publishing Group UK London}
}

@article{HeshenWang,
author = {Wang, Heshen and Xu, Runzhang and Liu, Cai and Wang, Le and Zhang, Zhan and Su, Huimin and Wang, Shanmin and Zhao, Yusheng and Liu, Zhaojun and Yu, Dapeng and Mei, Jia-Wei and Zou, Xiaolong and Dai, Jun-Feng},
title = {{Pressure-Dependent Intermediate Magnetic Phase in Thin $\mathrm{Fe}_3\mathrm{GeTe}_2$ Flakes}},
journal = {The Journal of Physical Chemistry Letters},
volume = {11},
number = {17},
pages = {7313-7319},
year = {2020},
doi = {10.1021/acs.jpclett.0c01801}
    

}

@article{XiangqiWang,
  title = {{Pressure-induced modification of the anomalous Hall effect in layered $\mathrm{Fe}_3\mathrm{GeTe}_2$}},
  author = {Wang, Xiangqi and Li, Zeyu and Zhang, Min and Hou, Tao and Zhao, Jinggeng and Li, Lin and Rahman, Azizur and Xu, Zilong and Gong, Junbo and Chi, Zhenhua and Dai, Rucheng and Wang, Zhongping and Qiao, Zhenhua and Zhang, Zengming},
  journal = {Physical Review B},
  volume = {100},
  issue = {1},
  pages = {014407},
  numpages = {6},
  year = {2019},
  month = {Jul},
  publisher = {American Physical Society},
  doi = {10.1103/PhysRevB.100.014407},
  url = {https://link.aps.org/doi/10.1103/PhysRevB.100.014407}
}

@article{RyujiFujita,
author = {Fujita, Ryuji and Gurung, Gautam and Mawass, Mohamad-Assaad and Smekhova, Alevtina and Kronast, Florian and Toh, Alexander Kang-Jun and Soumyanarayanan, Anjan and Ho, Pin and Singh, Angadjit and Heppell, Emily and Backes, Dirk and Maccherozzi, Francesco and Watanabe, Kenji and Taniguchi, Takashi and Mayoh, Daniel A. and Balakrishnan, Geetha and van der Laan, Gerrit and Hesjedal, Thorsten},
title = {{Strain-Modulated Ferromagnetism at an Intrinsic van der Waals Heterojunction}},
journal = {Advanced Functional Materials},
volume = {34},
number = {36},
pages = {2400552},
doi = {https://doi.org/10.1002/adfm.202400552},
year = {2024}
}

@article{ding2022tuning,
  title={{Tuning the density of zero-field skyrmions and imaging the spin configuration in a two-dimensional $\mathrm{Fe}_3\mathrm{GeTe}_2$ magnet}},
  author={Ding, Bei and Li, Xue and Li, Zefang and Xi, Xuekui and Yao, Yuan and Wang, Wenhong},
  journal={NPG Asia Materials},
  volume={14},
  number={1},
  pages={74},
  year={2022},
  doi = {https://doi.org/10.1038/s41427-022-00418-z},
  publisher={Springer Japan Tokyo}
}

@article{Kim_2019,
doi = {10.1088/1361-6528/ab0a37},
url = {https://doi.org/10.1088/1361-6528/ab0a37},
year = {2019},
month = {mar},
publisher = {IOP Publishing},
volume = {30},
number = {24},
pages = {245701},
author = {Kim, Dongseuk and Park, Sijin and Lee, Jinhwan and Yoon, Jungbum and Joo, Sungjung and Kim, Taeyueb and Min, Kil-joon and Park, Seung-Young and Kim, Changsoo and Moon, Kyoung-Woong and Lee, Changgu and Hong, Jisang and Hwang, Chanyong},
title = {{Antiferromagnetic coupling of van der Waals ferromagnetic $\mathrm{Fe}_3\mathrm{GeTe}_2$}},
journal = {Nanotechnology}
}

@article{Yi_2017,
doi = {10.1088/2053-1583/4/1/011005},
url = {https://doi.org/10.1088/2053-1583/4/1/011005},
year = {2016},
month = {nov},
publisher = {IOP Publishing},
volume = {4},
number = {1},
pages = {011005},
author = {Yi, Jieyu and Zhuang, Houlong and Zou, Qiang and Wu, Zhiming and Cao, Guixin and Tang, Siwei and Calder, S A and Kent, P R C and Mandrus, David and Gai, Zheng},
title = {{Competing antiferromagnetism in a quasi-$\mathrm{2D}$ itinerant ferromagnet: $\mathrm{Fe}_3\mathrm{GeTe}_2$}},
journal = {2D Materials}
}

@article{Zhen-Xiong,
  title = {{Magnetic ground state and electron-doping tuning of Curie temperature in $\mathrm{Fe}_3\mathrm{GeTe}_2$: First-principles studies}},
  author = {Shen, Zhen-Xiong and Bo, Xiangyan and Cao, Kun and Wan, Xiangang and He, Lixin},
  journal = {Physical Review B},
  volume = {103},
  issue = {8},
  pages = {085102},
  numpages = {6},
  year = {2021},
  month = {Feb},
  publisher = {American Physical Society},
  doi = {10.1103/PhysRevB.103.085102},
  url = {https://link.aps.org/doi/10.1103/PhysRevB.103.085102}
}

@article{PUSHKAREV2023171456,
title = {{An effective spin model on the honeycomb lattice for the description of magnetic properties in two-dimensional $\mathrm{Fe}_3\mathrm{GeTe}_2$}},
journal = {Journal of Magnetism and Magnetic Materials},
volume = {588},
pages = {171456},
year = {2023},
issn = {0304-8853},
doi = {https://doi.org/10.1016/j.jmmm.2023.171456},
url = {https://www.sciencedirect.com/science/article/pii/S030488532301106X},
author = {Georgy V. Pushkarev and Danis I. Badrtdinov and Ilia A. Iakovlev and Vladimir V. Mazurenko and Alexander N. Rudenko}
}

@article{siesta,
doi = {10.1088/0953-8984/14/11/302},
url = {https://doi.org/10.1088/0953-8984/14/11/302},
year = {2002},
month = {mar},
publisher = {},
volume = {14},
number = {11},
pages = {2745},
author = {José M Soler and Emilio Artacho and Julian D Gale and Alberto García and Javier Junquera and Pablo Ordejón and Daniel Sánchez-Portal},
title = {{The SIESTA method for $\textit{ab}$ $\textit{initio}$ order-$\mathrm{N}$
materials simulation}},
journal = {Journal of Physics: Condensed Matter}
}

@article{PBE,
  title = {{Generalized Gradient Approximation Made Simple}},
  author = {Perdew, John P. and Burke, Kieron and Ernzerhof, Matthias},
  journal = {Physical Review Letters},
  volume = {77},
  issue = {18},
  pages = {3865--3868},
  numpages = {0},
  year = {1996},
  month = {Oct},
  publisher = {American Physical Society},
  doi = {10.1103/PhysRevLett.77.3865},
  url = {https://link.aps.org/doi/10.1103/PhysRevLett.77.3865}
}

@article{Troullier,
  title = {{Efficient pseudopotentials for plane-wave calculations}},
  author = {Troullier, N. and Martins, Jos\'e Lu\'{\i}s},
  journal = {Physical Review B},
  volume = {43},
  issue = {3},
  pages = {1993--2006},
  numpages = {0},
  year = {1991},
  month = {Jan},
  publisher = {American Physical Society},
  doi = {10.1103/PhysRevB.43.1993},
  url = {https://link.aps.org/doi/10.1103/PhysRevB.43.1993}
}

@software{grogupy,
  author  = {Pozs{\'a}r, D{\'a}niel Tibor and Mart{\'i}nez-Carracedo, Gabriel and Garc{\'i}a-Fuente, Amador and Udvardi, L{\'a}szl{\'o} and Szunyogh, L{\'a}szl{\'o} and Ferrer, Jaime and Oroszl{\'a}ny, L{\'a}szl{\'o}},
  title   = {{grogupy}: v0.4.0},
  year    = {2025},
  doi     = {10.5281/zenodo.15449541},
  url     = {https://doi.org/10.5281/zenodo.15449541}
}

@article{LIECHTENSTEIN198765,
title = {{Local spin density functional approach to the theory of exchange interactions in ferromagnetic metals and alloys}},
journal = {Journal of Magnetism and Magnetic Materials},
volume = {67},
number = {1},
pages = {65-74},
year = {1987},
issn = {0304-8853},
doi = {https://doi.org/10.1016/0304-8853(87)90721-9},
url = {https://www.sciencedirect.com/science/article/pii/0304885387907219},
author = {A.I. Liechtenstein and M.I. Katsnelson and V.P. Antropov and V.A. Gubanov}
}

@article{LUdvardi,
  title = {{First-principles relativistic study of spin waves in thin magnetic films}},
  author = {Udvardi, L. and Szunyogh, L. and Palot\'as, K. and Weinberger, P.},
  journal = {Physical Review B},
  volume = {68},
  issue = {10},
  pages = {104436},
  numpages = {11},
  year = {2003},
  month = {Sep},
  publisher = {American Physical Society},
  doi = {10.1103/PhysRevB.68.104436},
  url = {https://link.aps.org/doi/10.1103/PhysRevB.68.104436}
}

@article{Houlong,
  title = {{Strong anisotropy and magnetostriction in the two-dimensional Stoner ferromagnet $\mathrm{Fe}_3\mathrm{GeTe}_2$}},
  author = {Zhuang, Houlong L. and Kent, P. R. C. and Hennig, Richard G.},
  journal = {Physical Review B},
  volume = {93},
  issue = {13},
  pages = {134407},
  numpages = {7},
  year = {2016},
  month = {Apr},
  publisher = {American Physical Society},
  doi = {10.1103/PhysRevB.93.134407},
  url = {https://link.aps.org/doi/10.1103/PhysRevB.93.134407}
}

@article{roemer2020robust,
  title={{Robust ferromagnetism in wafer-scale monolayer and multilayer $\mathrm{Fe}_3\mathrm{GeTe}_2$}},
  author={Roemer, Ryan and Liu, Chong and Zou, Ke},
  journal={npj 2D Materials and Applications},
  doi = {10.1038/s41699-020-00167-z},
  url = {https://doi.org/10.1038/s41699-020-00167-z},
  volume={4},
  number={1},
  pages={33},
  year={2020},
  publisher={Nature Publishing Group UK London}
}

@article{Landau:437299,
      author        = "Landau, Lev Davidovich and Lifshitz, E",
      title         = "{{On the theory of the dispersion of magnetic permeability
                       in ferromagnetic bodies}}",
      journal       = "Reprinted from Physikalische Zeitschrift der Sowjetunion",
      volume        = "8",
      pages         = "153",
      year          = "1935",
      doi = {https://doi.org/10.1016/B978-0-08-036364-6.50008-9},
       url = {https://www.sciencedirect.com/science/article/pii/B9780080363646500089}
}

@article{Nagyfalusi_2022,
doi = {10.1088/1361-648X/ac8260},
url = {https://doi.org/10.1088/1361-648X/ac8260},
year = {2022},
month = {jul},
publisher = {IOP Publishing},
volume = {34},
number = {39},
pages = {395803},
author = {Nagyfalusi, B and Udvardi, L and Szunyogh, L},
title = {{Magnetic ground state of supported monatomic $\mathrm{Fe}$ chains from first principles}},
journal = {Journal of Physics: Condensed Matter}
}

@article{Moriya,
  title = {{Anisotropic Superexchange Interaction and Weak Ferromagnetism}},
  author = {Moriya, T\^oru},
  journal = {Physical Review},
  volume = {120},
  issue = {1},
  pages = {91--98},
  numpages = {0},
  year = {1960},
  month = {Oct},
  publisher = {American Physical Society},
  doi = {10.1103/PhysRev.120.91},
  url = {https://link.aps.org/doi/10.1103/PhysRev.120.91}
}

@article{MossbauerSpectro,
author = {Dang, Ngoc-Toan and Kozlenko, Denis P. and Lis, Olga N. and Kichanov, Sergey E. and Lukin, Yevgenii V. and Golosova, Natalia O. and Savenko, Boris N. and Duong, Dinh-Loc and Phan, The-Long and Tran, Tuan-Anh and Phan, Manh-Huong},
title = {{High Pressure-Driven Magnetic Disorder and Structural Transformation in $\mathrm{Fe}_3\mathrm{GeTe}_2$: Emergence of a Magnetic Quantum Critical Point}},
journal = {Advanced Science},
volume = {10},
number = {9},
pages = {2206842},
doi = {https://doi.org/10.1002/advs.202206842},
year = {2023}
}

@article{Rozsa_2014,
doi = {10.1088/0953-8984/26/21/216003},
url = {https://doi.org/10.1088/0953-8984/26/21/216003},
year = {2014},
month = {may},
publisher = {IOP Publishing},
volume = {26},
number = {21},
pages = {216003},
author = {Rózsa, L and Udvardi, L and Szunyogh, L},
title = {{Langevin spin dynamics based on $\textit{ab}$ $\textit{initio}$ calculations: numerical schemes and applications}},
journal = {Journal of Physics: Condensed Matter}
}

@article{polak1969note,
     author = {Polak, E. and Ribiere, G.},
     title = {Note sur la convergence de m\'ethodes de directions conjugu\'ees},
     journal = {Revue fran\c{c}aise d'informatique et de recherche op\'erationnelle. S\'erie rouge},
     pages = {35--43},
     year = {1969},
     publisher = {Dunod},
     address = {Paris},
     volume = {3},
     number = {R1},
     mrnumber = {255025},
     zbl = {0174.48001},
     url = {https://www.numdam.org/item/M2AN_1969__3_1_35_0/}
}

@article{fei2018two,
  title={{Two-dimensional itinerant ferromagnetism in atomically thin $\mathrm{Fe}_3\mathrm{GeTe}_2$}},
  author={Fei, Zaiyao and Huang, Bevin and Malinowski, Paul and
            Wang, Wenbo and Song, Tiancheng and Sanchez, Joshua and
            Yao, Wang and Xiao, Di and Zhu, Xiaoyang and
            May, Andrew F. and Wu, Wei and Cobden, David H. and
            Chu, Jiun-Haw and Xu, Xiaodong},
  journal={Nature Materials},
  doi = {10.1038/s41563-018-0149-7},
  url = {https://doi.org/10.1038/s41563-018-0149-7},
  volume={17},
  number={9},
  pages={778--782},
  year={2018},
  publisher={Nature Publishing Group UK London}
}

@article{ghosh2023unraveling,
  title={{Unraveling effects of electron correlation in two-dimensional $\mathrm{Fe}_n\mathrm{GeTe}_2$ (n= 3, 4, 5) by dynamical mean field theory}},
  author={Ghosh, Sukanya and Ershadrad, Soheil and Borisov, Vladislav and Sanyal, Biplab},
  journal={npj Computational Materials},
  doi = {10.1038/s41524-023-01024-5},
  url = {https://doi.org/10.1038/s41524-023-01024-5},
  volume={9},
  number={1},
  pages={86},
  year={2023},
  publisher={Nature Publishing Group UK London}
}

@article{gabriel,
  title = {{Relativistic magnetic interactions from nonorthogonal basis sets}},
  author = {Mart\'{\i}nez-Carracedo, Gabriel and Oroszl\'any, L\'aszl\'o and Garc\'{\i}a-Fuente, Amador and Ny\'ari, Bendeg\'uz and Udvardi, L\'aszl\'o and Szunyogh, L\'aszl\'o and Ferrer, Jaime},
  journal = {Physical Review B},
  volume = {108},
  issue = {21},
  pages = {214418},
  numpages = {18},
  year = {2023},
  month = {Dec},
  publisher = {American Physical Society},
  doi = {10.1103/PhysRevB.108.214418},
  url = {https://link.aps.org/doi/10.1103/PhysRevB.108.214418}
}

@article{Laci-2019,
  title = {{Exchange interactions from a nonorthogonal basis set: From bulk ferromagnets to the magnetism in low-dimensional graphene systems}},
  author = {Oroszl\'any, L\'aszl\'o and Ferrer, Jaime and De\'ak, Andr\'as and Udvardi, L\'aszl\'o and Szunyogh, L\'aszl\'o},
  journal = {Physical Review B},
  volume = {99},
  issue = {22},
  pages = {224412},
  numpages = {12},
  year = {2019},
  month = {Jun},
  publisher = {American Physical Society},
  doi = {10.1103/PhysRevB.99.224412},
  url = {https://link.aps.org/doi/10.1103/PhysRevB.99.224412}
}

@inbook{Nowak2007,
	author = {Nowak, Ulrich},
	booktitle = {{Handbook of Magnetism and Advanced Magnetic Materials}},
	chapter = {},
	doi = {10.1002/9780470022184.hmm205},
	isbn = {9780470022184},
	pages = {},
	publisher = {John Wiley \& Sons, Ltd},
	title = {Classical Spin Models},
	url = {https://onlinelibrary.wiley.com/doi/abs/10.1002/9780470022184.hmm205},
	year = {2007}
}

@article{May2016,
  title = {{Magnetic structure and phase stability of the van der Waals bonded ferromagnet $\mathrm{Fe}_{3-x}\mathrm{GeTe}_2$}},
  author = {May, Andrew F. and Calder, Stuart and Cantoni, Claudia and Cao, Huibo and McGuire, Michael A.},
  journal = {Physical Review B},
  volume = {93},
  issue = {1},
  pages = {014411},
  numpages = {11},
  year = {2016},
  month = {Jan},
  publisher = {American Physical Society},
  doi = {10.1103/PhysRevB.93.014411},
  url = {https://link.aps.org/doi/10.1103/PhysRevB.93.014411}
}

@article{Rozsa2015,
  title = {{Magnetic phase diagram of an $\mathrm{Fe}$ monolayer on $\mathrm{W}(110)$ and $\mathrm{Ta}(110)$ surfaces based on $\textit{ab}$ $\textit{initio}$ calculations}},
  author = {R\'ozsa, Levente and Udvardi, L\'aszl\'o and Szunyogh, L\'aszl\'o and Szab\'o, Istv\'an A.},
  journal = {Physical Review B},
  volume = {91},
  issue = {14},
  pages = {144424},
  numpages = {13},
  year = {2015},
  month = {Apr},
  publisher = {American Physical Society},
  doi = {10.1103/PhysRevB.91.144424},
  url = {https://link.aps.org/doi/10.1103/PhysRevB.91.144424}
}

@article{Jenkins2022,
  title={{Breaking through the $\mathrm{Mermin}$-$\mathrm{Wagner}$ limit in $2\mathrm{D}$ van der Waals magnets}},
  author={Jenkins, Sarah and R{\'o}zsa, Levente and Atxitia, Unai and Evans, Richard FL and Novoselov, Kostya S and Santos, Elton JG},
  journal={Nature Communications},
  volume={13},
  number={1},
  pages={6917},
  year={2022},
  publisher={Nature Publishing Group UK London},
  url = {https://doi.org/10.1038/s41467-022-34389-0},
  DOI = {10.1038/s41467-022-34389-0}
}

@article{Laszloffy2019,
  title = {{Magnetic structure of monatomic $\mathrm{Fe}$ chains on $\mathrm{Re(0001)}$: Emergence of chiral multispin interactions}},
  author = {L\'aszl\'offy, A. and R\'ozsa, L. and Palot\'as, K. and Udvardi, L. and Szunyogh, L.},
  journal = {Physical Review B},
  volume = {99},
  issue = {18},
  pages = {184430},
  numpages = {13},
  year = {2019},
  month = {May},
  publisher = {American Physical Society},
  doi = {10.1103/PhysRevB.99.184430},
  url = {https://link.aps.org/doi/10.1103/PhysRevB.99.184430}
}

@article{Cuadrado_2012,
doi = {10.1088/0953-8984/24/8/086005},
url = {https://doi.org/10.1088/0953-8984/24/8/086005},
year = {2012},
month = {jan},
publisher = {IOP Publishing},
volume = {24},
number = {8},
pages = {086005},
author = {Cuadrado, R and Cerdá, J I},
title = {{Fully relativistic pseudopotential formalism under an atomic orbital basis: spin–orbit splittings and magnetic anisotropies}},
journal = {Journal of Physics: Condensed Matter}
}

@dataset{jyothi_bhasu_2026_20737279,
  author       = {Jyothi Bhasu, Anjali and
                  Szunyogh, Laszlo},
  title        = {{Dataset for "Strain induced magnetic phase
                   transitions in $\mathrm{Fe}_3\mathrm{GeTe}_2$ monolayer"
                  }},
  month        = jun,
  year         = 2026,
  publisher    = {Zenodo},
  version      = {v1},
  doi          = {10.5281/zenodo.20737279},
  url          = {https://doi.org/10.5281/zenodo.20737279},
}

\end{document}